%% file: 3d-nmpc-paper.tex
\documentclass[letterpaper, 10 pt, conference]{ieeeconf}  

\IEEEoverridecommandlockouts                              

\usepackage{amssymb}
\usepackage{amsmath}
\usepackage{graphicx}
\usepackage{bm}
\usepackage{hyperref}
\usepackage{comment}
\usepackage[nolist,nohyperlinks]{acronym}
\usepackage{tabularx}
\usepackage{subcaption}

\usepackage{tikz}

\usepackage{tikz-3dplot}
\usepackage{tikzscale}
\usetikzlibrary{calc}
\usetikzlibrary{shapes.geometric}

\usepackage{cancel}

\newcommand{\neglectedblock}[1]{%
  \begingroup
  \setlength{\fboxsep}{1pt}%
  \mathchoice
    {\colorbox{gray!18}{$\displaystyle #1$}}
    {\colorbox{gray!18}{$\textstyle #1$}}
    {\colorbox{gray!18}{$\scriptstyle #1$}}
    {\colorbox{gray!18}{$\scriptscriptstyle #1$}}
  \endgroup
}

\usepackage{pgfplots}
\usepackage{pgfplotstable}
\pgfplotsset{compat=1.18}
\usepgfplotslibrary{groupplots}
\RequirePackage[pgfplots,svgnames,x11names]{tumcolor}%
\pgfplotsset{
  colormap={TUMBlueMap}{
    rgb(0pt)=(0.0,0.2,0.35)        
    rgb(25pt)=(0.0,0.32,0.58)      
    rgb(50pt)=(0.0,0.4,0.74)       
    rgb(75pt)=(0.39,0.63,0.78)     
    rgb(100pt)=(0.6,0.78,0.92)     
  }
}
\pgfplotsset{
  colormap={TUMBlueOrange}{
    rgb(0pt)=(0.0,0.2,0.35)        
    rgb(25pt)=(0.0,0.4,0.74)       
    rgb(50pt)=(0.6,0.78,0.92)     
    rgb(75pt)=(0.89,0.45,0.13)     
    rgb(100pt)=(0.45,0.23,0.07)     
  }
}

\usepackage{algorithm}
\usepackage{algpseudocode}
\usepackage{soul} 
\usepackage{siunitx}
\usepackage{mathtools}

\usetikzlibrary{external}
\usepackage[autostyle]{csquotes}
\MakeAutoQuote{‘}{’} 

\colorlet{Dynamic3d}{TUMBlue}%
\colorlet{Static3d}{TUMOrange!110}%
\colorlet{Plane2d}{TUMGrayLighter!65}%

\usepackage[
    backend=bibtex, 
    doi=false,
    isbn=false,
    url=false,          
    eprint=false, 
    sorting=none,
    maxbibnames=2,      
    minbibnames=1       
]{biblatex}

\graphicspath{{figures/}}

\AtBeginBibliography{\small}

\input{CustomCommands.tex}

\input{acronyms}

\usepackage{fancyhdr}
\usepackage{etoolbox}

\newcommand\copyrighttext{%
	\footnotesize \textcopyright 2026 EUCA. Accepted for publication at the 24th European Control Conference (ECC), Reykjavík, Iceland.
}
\makeatletter
\patchcmd{\@maketitle}{\thispagestyle{plain}}{\thispagestyle{firstpage}}{}{}
\makeatother

\fancypagestyle{firstpage}{
	\fancyhf{} 
	\fancyfoot[C]{%
		\fbox{%
			\parbox{\dimexpr\textwidth-2\fboxsep-2\fboxrule\relax}{%
				\centering \copyrighttext
			}%
		}%
	}

}

\title{\LARGE \bf
Robust Nonlinear Trajectory Tracking Control for Autonomous Racing on Three-Dimensional Tracks
}

\author{Joscha F. Bongard$^{1, 3, *}$, Georg Jank$^{2,3, *}$, Simon Sagmeister$^{1,4}$, Boris Lohmann$^{3}$
\thanks{* Authors have contributed equally, and names are in alphabetical order.}
\thanks{$^{1}$ funded by the Deutsche Forschungsgemeinschaft (DFG, German Research Foundation) - 469341384}
\thanks{$^{2}$ Corresponding Author}
\thanks{$^{3}$ Chair of Automatic Control, Department of Engineering Physics and Computation, Technical University of Munich,
        Boltzmannstraße 15, 85748 Garching bei München, Germany
        {\tt\small \{joscha.bongard,georg.jank,lohmann\}@tum.de}}%
\thanks{$^{4}$ Chair of Automotive Technology, Department of Engineering Physics and Computation, Technical University of Munich,
        Boltzmannstraße 15, 85748 Garching bei München, Germany
        {\tt\small simon.sagmeister@tum.de}}%
}
\begin{document}

\maketitle
\thispagestyle{firstpage}

\begin{abstract}
    We propose a robust nonlinear model predictive control (MPC) scheme for trajectory-tracking control of autonomous vehicles at the limits of handling on non-planar road surfaces. We derive the dynamics from first principles and selectively omit terms with negligible dynamic influence to maintain real-time capability. The resulting MPC with a three-dimensional (3D) dynamic single-track model integrates relevant dynamic effects directly into the prediction model and leverages them to improve prediction accuracy and therefore control performance. Even if the influence of terrain-induced vertical loads on the total acceleration potential is modeled, tire-road interactions are subject to uncertainty and disturbance. The uncertainty-aware constraint tightening scheme introduces a margin to constraint bounds to keep the vehicle controllable and stable in this environment. To validate our proposed approach, we perform high-fidelity dynamic double-track vehicle dynamics simulations on a model of a real circuit. We find that our algorithm can improve trajectory-tracking accuracy while maintaining low computation times.
\end{abstract}

\input{sections/preliminaries.tex}

\section{Methodology}

\input{sections/vehicle_model}

\input{sections/control_design.tex}

\input{sections/results.tex}

\input{sections/discussion.tex}

\section{Acknowledgments}
\emph{Author Contributions:}
Joscha F. Bongard: Conceptualization, Methodology, Software, Writing – original draft. Georg Jank: Conceptualization, Methodology, Software, Visualization, Writing – original draft. Simon Sagmeister: Software, Methodology, Writing – review \& editing. Boris Lohmann: Supervision, Writing – review \& editing, Funding acquisition, Project administration.



\printbibliography

\end{document}

%% file: CustomCommands.tex
\definecolor{tum blue}{HTML}{0065BD}
\definecolor{tum blue 1}{HTML}{98C6EA}
\definecolor{tum blue 2}{HTML}{64A0C8}
\definecolor{tum blue 3}{HTML}{0073CF}
\definecolor{tum blue 4}{HTML}{005293}
\definecolor{tum blue 5}{HTML}{003359}

\definecolor{tum green}{HTML}{A2AD00}
\definecolor{tum orange}{HTML}{E37222}
\definecolor{tum ivory}{HTML}{DAD7CB}

\definecolor{tum dia violet}{HTML}{69085A}
\definecolor{tum dia dark blue}{HTML}{0F1B5F}
\definecolor{tum dia turquoise}{HTML}{00778A}
\definecolor{tum dia dark green}{HTML}{007C30}
\definecolor{tum dia light green}{HTML}{679A1D}
\definecolor{tum dia light yellow}{HTML}{FFDC00}
\definecolor{tum dia dark yellow}{HTML}{F9BA00}
\definecolor{tum dia dark orange}{HTML}{D64C13}
\definecolor{tum dia red}{HTML}{C4071B}
\definecolor{tum dia dark red}{HTML}{9C0D16}

\newcommand{\R}{\mathbb{R}}



%% file: acronyms.tex
\acrodef{mpc}[MPC]{Model Predictive Control}
\acrodef{3d}[3D]{Three-Dimensional}
\acrodef{2d}[2D]{Two-Dimensional}
\acrodef{ct}[CT]{Constraint Tightening}
\acrodef{mf}[MF]{Magic Formula}
\acrodef{dstm}[DSTM]{Dynamic Single Track Model}
\acrodef{cog}[COG]{Center of Gravity}
\acrodef{ocp}[OCP]{Optimal Control Problem}
\acrodef{ccm}[CCM]{Control Contraction Metric}
\acrodef{ad}[AD]{Autonomous Driving}
\acrodef{nlp}[NLP]{Nonlinear Program}
\acrodef{sqp}[SQP]{Sequential Quadratic Program}
\acrodef{qp}[QP]{Quadratic Program}
\acrodef{rti}[RTI]{Real-Time Iteration}

%% file: sections/preliminaries.tex
\section{Preliminaries}

\subsection{Introduction}

\begin{figure}
    \centering
    \includegraphics[width=\linewidth]{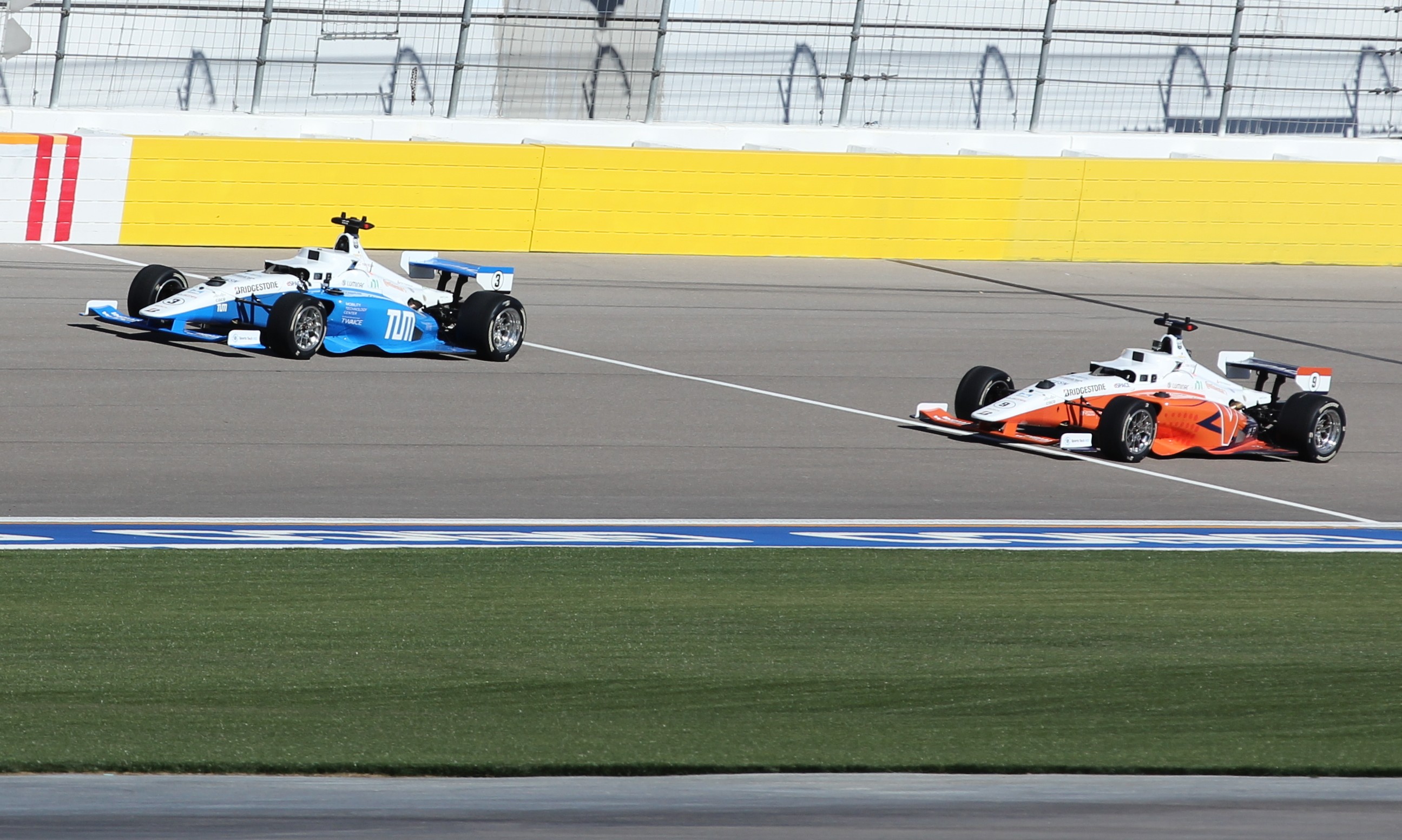}
    \caption{Autonomous vehicles \emph{Dallara AV24} of \emph{TUM Autonomous Motorsport} and \emph{Unimore} in a banked turn at the \emph{IAC}.}
    \label{fig:iac}
\end{figure}

Autonomous racing series like the \emph{A2RL} and the \emph{IAC} (see Figure~\ref{fig:iac}) push algorithms for \ac{ad} to the limits in pursuit of lap times and competitive racing maneuvers as shown by expert drivers~\cite{hoffmann2025a2rl, Betz2022TUMAutonomousMotorsport}. 
This paper focuses on the trajectory tracking module, developed to accurately follow a pre-planned path and velocity while maintaining vehicle stability and adherence to path and dynamics constraints.
\acf{mpc} offers a powerful, optimization-based control framework that balances competing objectives and handles coupled constraints typical in \ac{ad} applications.
However, the effectiveness of \ac{mpc} relies on the accuracy of the prediction model in determining how current and future inputs influence future states.
Model mismatch degrades performance and can result in a loss of closed-loop stability~\cite{Rawlings2020MPC}. While a \ac{ct} can help to avoid constraint violations from model mismatch, it is usually only feasible for simple models that offer poor predictive quality at the limits~\cite{Stano2023}.
While the effects of the \ac{3d} geometry of road surfaces are often neglected in the design of tracking control modules, their explicit consideration has proven advantageous in related \ac{ad} modules, such as trajectory planning~\cite{Rowold2023} and state estimation~\cite{Goblirsch2024}.
Therefore, this paper studies the potential benefits of integrating dynamic \ac{3d} effects in the prediction model for robust \ac{mpc}.

\subsection{Related Work}
\ac{mpc} has emerged as the de facto standard for high-performance trajectory tracking in \ac{ad} applications, particularly racing, offering a flexible framework for approximately optimal control of nonlinear dynamics under constraints \cite{Rawlings2020MPC,Stano2023,Betz2022}.

The prediction model is one of the fundamental components defining the behavior of an \ac{mpc} \cite{Schwenzer2021}.
Linear time invariant \cite{Wischnewski2021TubeMPC}, linear time-varying \cite{Wang2021}, and linear parameter-varying \cite{Alcala2020} prediction models are computationally light and can be simple to parametrize. Conversely, nonlinear models allow utilization of more dynamic potential through better predictive quality at the dynamic limits~\cite{Raji2022,Bongard2025IVRMPC}. While more computationally demanding, \ac{mpc} schemes based on the \ac{dstm} have become widely adopted for \ac{ad} in recent years for driving at the dynamic limits due to their accurate representation of tire and vehicle nonlinearities~\cite{Stano2023}. While several works have demonstrated the effectiveness of this model formulation in high-speed racing, most rely on simplified planar models that neglect the influence of 3D racetrack geometry entirely~\cite{Bongard2025IVRMPC} or only treat banking and slope as static disturbances~\cite{Raji2022}.

Recent works in the adjacent field of raceline planning have shown that explicitly considering the accelerations from the rotating vehicle reference frame on the uneven racetrack can reduce lap times by making the prediction of vertical forces and therefore acceleration potential in turns more accurate \cite{Rowold2023, Lovato2022, Limebeer2015}. A similar improvement was observed for the state estimation of autonomous race cars, where inclusion of the rotating reference frame in the dynamics of the Extended Kalman Filter led to a significant reduction in positional error on banked oval tracks~\cite{Goblirsch2024}. 

Despite these advances, many \ac{mpc} designs for \ac{ad} show performance degradation on \ac{3d} racetracks by neglecting the dynamic effects of \ac{3d} road geometries in their prediction models.
Some \ac{mpc} schemes account for roll dynamics on banked surfaces~\cite{Liu2018} or roll- and pitch from compliant suspension on planar surfaces~\cite{Taghavifar2019}. However, terrain-induced combined roll- and pitch dynamics are mostly unexplored in \ac{mpc} plant models for autonomous racing and \ac{ad} in general.

\ac{ct} is a common method to enhance the robustness of constraint satisfaction under model mismatch by forcing a predictive back-off from constraint bounds. While this back-off is necessary for safety, it is detrimental to \ac{ad} tracking performance by reducing the available path or acceleration potential. The closer the prediction model is to the real vehicle, the smaller the back-off can be.\\
Tightening constraints by propagating uncertainties through prediction models is a challenging task to do non-conservatively and requires approximations, mostly explored for linear time-invariant models \cite{Wischnewski2022TubeMPCApproachHighSpeedOvals}.
A computationally efficient uncertainty-aware \ac{ct} was previously proposed for a purely \ac{2d} prediction, but its impact on \ac{3d} schemes was not explored~\cite{Bongard2025IVRMPC}.

\subsection{Contribution}
This paper presents a robust \ac{mpc} scheme that exploits \ac{3d} racetrack geometry to improve stability and tracking performance. Compared to previous work \cite{Bongard2025IVRMPC}, the main contributions are:
\begin{itemize}
    \item \emph{Dynamic \ac{3d} model:} We combine uncertainty-aware \ac{ct} with an extended \ac{dstm} defined in a rotating reference frame, enabling prediction of terrain-induced dynamics.
    \item \emph{Tire force modeling:} We extend the tire model to a load-dependent combined Pacejka formulation to capture terrain-induced dynamics and longitudinal load-transfer.
    \item \emph{System-level validation:} We implement the controller within a \texttt{C++}/ROS2 autonomous racing software stack and benchmark it in high-fidelity simulation against two relevant baselines, demonstrating improved tracking accuracy and robustness on a \ac{3d} racetrack.
\end{itemize}

\subsection{Notation}
We write derivatives of $x$ in time $t$ and path progress $s$ as $\dot{x}=\frac{\mathrm{d}x}{\mathrm{d}t}$ and $x'=\frac{\mathrm{d}x}{\mathrm{d}s}$. Prefix subscripts indicate the reference frame, e.\,g., $_\mathcal{A}\bm{\omega}_\mathcal{BC}$ is the angular velocity of $\mathcal{B}$ relative to $\mathcal{C}$ denoted in $\mathcal{A}$. We abbreviate $\sin{x}$ by $\mathrm{s}_x$ (not to be confused with progress $s$) and $\cos{x}$ by $\mathrm{c}_x$ where appropriate. For discrete-time dynamics, the signal $x$ at prediction step $k$ is written as $x_k$. $\mathbb{I}_{\left[a,\, b\right]}$ is the set of integers in the closed interval between $a$ and $b$.

%% file: sections/vehicle_model.tex
\subsection{Vehicle Model}\label{sec:vehicle_model}

The prediction model extends the \acf{dstm} of \cite{Raji2022,Bongard2025IVRMPC} with \ac{3d} racetrack rotations and a simplified load-dependent combined-slip Pacejka \ac{mf} tire model.
The \ac{dstm} lumps the wheels of each axle into one virtual wheel. 
Compared to the double-track vehicle model, with roll dynamics and individual tires, this simplified model reduces complexity while retaining the dominant lateral and longitudinal dynamics.
\begin{figure}
    \vspace{5mm}
    \centering
    \def\svgwidth{0.95\linewidth}
    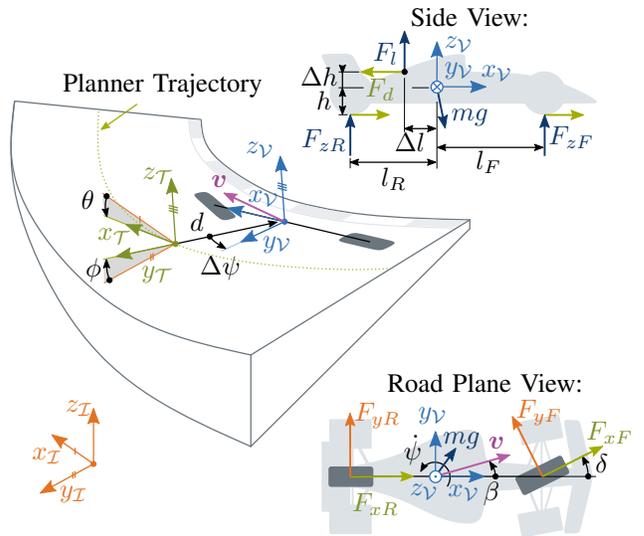
    \caption{Vehicle on banked and slope racetrack with relevant frames: Inertial frame $\mathcal{I}$, trajectory frame $\mathcal{T}$, and vehicle frame $\mathcal{V}$. 
    Note, the path deviation $d$ is shown negative (positive is to the left of the reference line).}
    \label{fig:3d-DSTM}
\end{figure}
\subsubsection{3D-DSTM}

To describe the motion of the vehicle on a \ac{3d} surface, we employ three reference frames, shown in \figureautorefname~\ref{fig:3d-DSTM}: 
the inertial frame $\mathcal{I}$, the trajectory frame $\mathcal{T}$, and the body-fixed vehicle frame $\mathcal{V}$.

$\mathcal{T}$ is defined with the x-axis $x_\mathcal{T}$ in the direction of the planner reference velocity $v_\mathrm{ref}$ and the z-axis $z_\mathcal{T}$ facing upwards, perpendicular to the road plane. 
$\mathcal{V}$ is fixed to the \ac{cog} of the vehicle with $x_\mathcal{V}$ pointing towards the center of the front axle, and $z_\mathcal{V}$ parallel to $z_\mathcal{T}$.

\paragraph{Relative kinematics}
The orientation of $\mathcal{T}$ relative to $\mathcal{I}$ is defined by three angles: heading $\chi$ (yaw), slope $\theta$ (pitch), and banking $\phi$ (roll). 
A position vector in the inertial frame $\mathcal{I}$ is transformed to the trajectory frame via:
\begin{equation}
    {_\mathcal{T}\bm{r}} = \mathbf{R}_x(\phi)\mathbf{R}_y(\theta)\mathbf{R}_z(\chi)\,{_\mathcal{I}\bm{r}} + {_\mathcal{T}\bm{r}_\mathcal{IT}},
\end{equation}
where $\mathbf{R}_x(\cdot)$, $\mathbf{R}_y(\cdot)$, $\mathbf{R}_z(\cdot)$ are rotation matrices about the $x$, $y$, $z$ axes respectively, and ${_\mathcal{T}\bm{r}_\mathcal{IT}}$ is the translation offset of the trajectory frame origin in $\mathcal{I}$.
The rotation rate over path progress $_\mathcal{T}\bm{\Omega}_\mathcal{TI}$ can be evaluated in the $\mathcal{T}$-frame as 
\begin{align}
_\mathcal{T}\bm{\Omega}_\mathcal{TI}&=
\begin{bmatrix} \hat{\phi}' & \hat{\theta}' & \hat{\chi}' \end{bmatrix}^\top\nonumber\\
&=\begin{bmatrix} \phi' \\ 0 \\ 0 \end{bmatrix} + 
\mathbf{R}_x(\phi)
\left(
\begin{bmatrix} 0 \\ \theta' \\ 0 \end{bmatrix} + 
\mathbf{R}_y(\theta) \begin{bmatrix} 0 \\ 0 \\ \chi' \end{bmatrix}
\right).\label{eq:omega_V}
\end{align}
This is derived from the three successive frame rotations: yaw ($\chi'$ in $z$), pitch ($\theta'$ in $y$), and then roll ($\phi'$ in $x$).
The values $\phi, \theta, \chi, \hat{\phi}', \hat{\theta}', \hat{\chi}'$ are properties of the reference trajectory, provided by the planning module \cite{Rowold2023}.

The vehicle $\mathcal{V}$ is positioned at a lateral offset $d$ from the trajectory reference line (positive to the left) and a heading offset $\Delta\psi$ from the racetrack heading. 
Thus, the angular velocity of the vehicle in the inertial frame is:
\begin{equation}
    {_\mathcal{V}\bm{\omega}_\mathcal{VI}} = {_\mathcal{V}\bm{\omega}_\mathcal{VT}} + \mathbf{R}_z(\Delta\psi)\,{_\mathcal{T}\bm{\Omega}_\mathcal{TI}}\dot{s},
\end{equation}
where $\dot{s}$ is the rate of progress along the path. 
For small heading deviations ($\Delta\psi \ll 1$) and using $\dot{\psi} = \Delta\dot{\psi} + \hat{\chi}'\dot{s}$, this simplifies to:
\begin{equation}
    {_\mathcal{V}\bm{\omega}_\mathcal{VI}} \approx \begin{bmatrix}\hat{\phi}'\dot{s} & \hat{\theta}'\dot{s} & \dot{\psi}\end{bmatrix}^\top.
\end{equation}
For compact notation we define $_\mathcal{V}\bm{\omega}_\mathcal{VI}=\begin{bmatrix}\hat{\omega}_x&\hat{\omega}_y&\hat{\omega}_z\end{bmatrix}^\top$.

\paragraph{Rigid-body dynamics}
The law of conservation of angular and linear momentum applied to the system yields
\begin{align}
        {_\mathcal{V}\mathbf{I}}\,{_\mathcal{V}\dot{\bm{\omega}}_\mathcal{VI}}+{_\mathcal{V}\bm{\omega}_\mathcal{VI}}\times({_\mathcal{V}\mathbf{I}}\,{_\mathcal{V}\bm{\omega}_\mathcal{VI}})&= \sum_i M_i \label{eq:DS1}\\
        m (_\mathcal{V}\dot{\bm{v}}+{_\mathcal{V}\bm{\omega}_\mathcal{VI}}\times{_\mathcal{V}\bm{v}})&=\sum_i F_i \label{eq:SPS1}
\end{align}
with vehicle moment of inertia $_\mathcal{V}\mathbf{I}=\mathrm{diag}(0,I_y,I_z)$, mass $m$ and velocities $v_x,v_y,v_z$ of the \ac{cog} projected in the road plane, i.\,e. 
$_\mathcal{V}\bm{v}=\begin{bmatrix}v_x+\hat{\omega}_yh&v_y-\hat{\omega}_xh&v_z\end{bmatrix}^\top$. 
We neglect roll dynamics and, therefore, the rolling moment of inertia. 
Substituting \eqref{eq:omega_V} into \eqref{eq:DS1} and evaluating the torque balance with the forces from \figureautorefname~\ref{fig:3d-DSTM}, we obtain
\begin{equation}
\begin{aligned}
    &\begin{bmatrix}
        (I_z-I_y)\hat{\omega}_z\hat{\omega}_y\\
        \neglectedblock{I_y\dot{\hat{\omega}}_y}-I_z\hat{\omega}_z\hat{\omega}_x\\
        I_z\dot{\hat{\omega}}_z+\neglectedblock{I_y\hat{\omega}_x\hat{\omega}_y}
    \end{bmatrix}
    = \begin{bmatrix}-l_R\\0\\-h\end{bmatrix}\times\begin{bmatrix}F_{xR}\\F_{yR}\\F_{zR}\end{bmatrix}\\
    &+ \begin{bmatrix}l_F\\0\\-h\end{bmatrix}\times\begin{bmatrix}F_{xF}\mathrm{c}_\delta-F_{yF}\mathrm{s}_\delta\\F_{yF}\mathrm{c}_\delta+F_{xF}\mathrm{s}_\delta\\F_{zF}\end{bmatrix}
    + \begin{bmatrix}-\Delta l\\0\\\Delta h\end{bmatrix}\times\begin{bmatrix}-F_{d}\\0\\F_{l}\end{bmatrix}.
\end{aligned}
\end{equation}
Evaluating the force balance \eqref{eq:SPS1} yields
\begin{subequations}
\begin{align}
    m\begin{bmatrix}\hat{a}_x\\\hat{a}_y\\\hat{a}_z\end{bmatrix} &=
    \begin{bmatrix}
        m(\dot{v}_x+\neglectedblock{\dot{\hat{\omega}}_yh}+{\hat{\omega}_yv_z}-\hat{\omega}_zv_y)\\
        m(\dot{v}_y-\neglectedblock{\dot{\hat{\omega}}_xh}+\hat{\omega}_zv_x-\hat{\omega}_xv_z)\\
        m(\dot{v}_z+\hat{\omega}_xv_y-\hat{\omega}_yv_x)
    \end{bmatrix}
    \\
    \begin{split}
    &= \begin{bmatrix}F_{xR}\\F_{yR}\\F_{zR}\end{bmatrix}
    + \begin{bmatrix}F_{xF}\mathrm{c}_\delta-F_{yF}\mathrm{s}_\delta\\F_{yF}\mathrm{c}_\delta+F_{xF}\mathrm{s}_\delta\\F_{zF}\end{bmatrix}
    + \begin{bmatrix}-F_{d}\\0\\F_{l}\end{bmatrix}\\
    &\quad+ \mathbf{R}_z(\Delta\psi)\mathbf{R}_x(\phi)\mathbf{R}_y(\theta)\mathbf{R}_z(\chi)(-mg)\bm{e}_z.
    \end{split}
\end{align}
\end{subequations}

Based on \tableautorefname~\ref{tab:orders_of_magnitude}, we neglect the gray-marked terms in \eqref{eq:DS1} and \eqref{eq:SPS1}. 
These are approximately three to four orders smaller than the dominant terms over the considered operating range. 
Hence, removing these terms has a minor impact on the dynamics while reducing model complexity.

\begin{table}[htbp]
    \vspace{5mm}
    \centering
    \begin{tabularx}{\linewidth}{c|c}
    \hline
    Quantities & Order of Magnitude\\
    \hline
    $\hat{\omega}_x, \hat{\omega}_y, \dot{\hat{\omega}}_x, \dot{\hat{\omega}}_y, h, \Delta h, \Delta l, \delta, \Delta \psi, \theta, \phi$ & $10^{-1}$\\
    $\hat{\omega}_z, \dot{\hat{\omega}}_z, v_y, v_z, l_F, l_R,d,\dot{d}$ & $10^0$\\
    $v_x,I_y$ & $10^2$\\
    $m,I_z,F_l,F_d$ & $10^3$\\
    $F_{xF},F_{xR},F_{yF},F_{yR},F_{zF},F_{zR}$ & $10^4$\\
    \hline
    \end{tabularx}
    \begin{tabularx}{\linewidth}{c|c|c|c}
         \multicolumn{4}{c}{Conservation of angular momentum \eqref{eq:DS1}} \\
         \hline 
         Term & OoM in \linebreak $\mathrm{N\,m}$ & Term & OoM in \linebreak $\mathrm{N\,m}$\\
         \hline
         $I_y\dot{\hat{\omega}}_y$ & $10^1$  & $I_z\dot{\hat{\omega}}_z$ & $10^3$ \\
         $I_z\hat{\omega}_z\hat{\omega}_x$ & $10^2$ & $I_y\hat{\omega}_x\hat{\omega}_y$ & $10^0$ \\
         $l_{\{F,R\}} F_{z,\{F,R\}}$ & $10^4$ & $l_{\{F,R\}} F_{y,\{F,R\}}$ & $10^4$ \\
         $h F_{x,\{F,R\}}$ & $10^3$   & $\Delta l F_{l},\Delta h F_{d}$ & $10^2$
    \end{tabularx}
        \begin{tabularx}{\linewidth}{c|c|c|c}
        \hline
         \multicolumn{4}{c}{Conservation of linear momentum \eqref{eq:SPS1}} \\
         \hline 
         Term & OoM in $\mathrm{N}$ & Term & OoM in $\mathrm{N}$\\
         \hline
         $m\dot{v}_x$ & $10^4$ & $m\dot{v}_y$ & $10^4$\\
         $m\dot{v}_z=m(\dot{\hat{\omega}}_xd+\hat{\omega}_x\dot{d})$ & $10^2$ & $m\dot{\hat{\omega}}_yh$ & $10^1$\\
         $m\dot{\hat{\omega}}_xh$&$10^1$ & $m\hat{\omega}_yv_z$&$10^2$ \\
         $m\hat{\omega}_zv_x$ & $10^5$ & $m\hat{\omega}_xv_y$ & $10^2$\\
         $m\hat{\omega}_zv_y$& $10^3$ & $m\hat{\omega}_xv_z$& $10^2$\\
         $m\hat{\omega}_yv_x$ & $10^4$ & $mg\mathrm{c}_\theta \mathrm{s}_{\Delta\psi}$ & $10^3$\\
         $mg\mathrm{c}_\theta \mathrm{s}_\phi \mathrm{s}_{\Delta\psi}$ & $10^2$& $mg\mathrm{s}_\theta \mathrm{s}_{\Delta\psi}$ & $10^2$\\
         $mg\mathrm{c}_\theta \mathrm{s}_\phi \mathrm{c}_{\Delta\psi}$ & $10^3$ & $mg\mathrm{c}_\theta \mathrm{c}_\phi$&$10^4$\\
         \hline
    \end{tabularx}
    \caption{Order of magnitude of terms in \eqref{eq:DS1} and \eqref{eq:SPS1}}
    \label{tab:orders_of_magnitude}
\end{table}

\paragraph{Equations of motion}
The equations of motion are given by the state and input vectors
\begin{subequations}
\begin{align}
    \bm{x}&=\begin{bmatrix}d&\Delta\psi&v_x&v_y&\dot{\psi}&\delta&T&B\end{bmatrix}^\top\label{eq:state_vector},\\
    \bm{u}&=\begin{bmatrix}u_{\mathrm{d}\delta}&u_{\mathrm{d}T}&u_{\mathrm{d}B}\end{bmatrix}^\top, \label{eq:input_vector}
\end{align}
\end{subequations}
and state dynamics
\begin{subequations}
\begin{align} \label{eq:dstm_dynamics}
    \dot{d} &= v_x \mathrm{s}_{\Delta\psi} + v_y \mathrm{c}_{\Delta\psi} , \\
    \Delta\dot{\psi} &= \dot{\psi} - \hat{\chi}' \dot{s} , \\
    \dot{v}_x &= \tilde{a}_x +g \mathrm{s}_\theta \mathrm{c}_{\Delta\psi} - g\mathrm{c}_\theta \mathrm{s}_\phi \mathrm{s}_{\Delta\psi}- \hat{\theta}' \dot{s} v_z + \dot{\psi} v_y , \\
    \dot{v}_y &= \tilde{a}_y -g \mathrm{s}_\theta \mathrm{s}_{\Delta\psi} - g \mathrm{c}_\theta \mathrm{s}_\phi \mathrm{c}_{\Delta\psi}-\dot{\psi}v_x+\hat{\phi}'\dot{s}v_z , \\
    \ddot{\psi} &= \frac{1}{I_z}(l_F F_{yF}\mathrm{c}_\delta+l_F F_{xF}\mathrm{s}_\delta-l_R F_{yR}),\\
    \dot{\delta} &= u_{\mathrm{d}\delta}, \hspace{0.2cm}
    \dot{T} = u_{\mathrm{d}T}, \hspace{0.2cm}
    \dot{B} = u_{\mathrm{d}B},
\end{align}
\end{subequations}
where $\delta$ is the steering angle, $T \in [0,1]$ is the normalized throttle, and $B \in [0,1]$ is the normalized brake. 
The progress rate $\dot{s}$ is approximated as $\dot{s}\approx\frac{v_x\mathrm{c}_{\Delta\psi}-v_y\mathrm{s}_{\Delta\psi}}{1-d\hat{\chi}'}$, which follows from Frenet-frame kinematics: the numerator is the vehicle velocity projected onto the trajectory tangent, while the denominator accounts for the vehicle position relative to path curvature. 
$v_z$ is given by $v_z\approx d\hat{\phi}'\dot{s}$, as $\mathcal{V}$ is offset by $d$ in the $\phi$-banked $\mathcal{T}$-frame. 
$F_l$ and $F_d$ are aerodynamic lift and drag forces, respectively, that are quadratically proportional to $v_x$.
The apparent accelerations in the body frame are given by
\begin{subequations}
\begin{align} \label{eq:accels_tilde}
	\tilde{a}_x &= \frac{1}{m}\left(F_{xF} \mathrm{c}_\delta - F_{yF} \mathrm{s}_\delta + F_{xR} - F_d\right),\\
	\tilde{a}_y &= \frac{1}{m}\left(F_{yF}\mathrm{c}_\delta+F_{xF}\mathrm{s}_\delta+F_{yR}\right).
\end{align}
\end{subequations}
The longitudinal $F_{xF},F_{xR}$, lateral $F_{yF},F_{yR}$, and vertical $F_{zF},F_{zR}$ tire forces depend on the vehicle state $\bm{x}$, orientation, and rotation on the uneven racetrack, as outlined in \sectionautorefname~\ref{sec:tire_forces}.

\subsubsection{Tire forces} In the following, we describe the calculation of longitudinal and vertical tire loads, and how they affect the lateral traction force via a load-dependent slip angle formulation of the Pacejka \ac{mf} tire model.
\label{sec:tire_forces}
\paragraph{Longitudinal and vertical load}
To represent rear-wheel drive with braking on both axles, we utilize normalized throttle and brake inputs $T, B \in [0,1]$ that determine the longitudinal tire forces on the front and rear axles via the throttle $C_T$, brake $C_{BF},C_{BR}$, and rolling resistance $C_{rr}$ constants \cite{Raji2022}
\begin{align}
    F_{xF}&=-C_{BF}B-\frac{mgC_{rr}l_R}{l_\mathrm{wb}},\\
    F_{xR}&=C_TT-C_{BR}B-\frac{mgC_{rr}l_F}{l_\mathrm{wb}},
\end{align}
with wheelbase $l_\mathrm{wb}=l_F+l_R$.\\
The vertical axle load is given by the balance of angular momentum in $y_\mathcal{V}$ and the balance of linear momentum in $z_\mathcal{V}$ with the assumption $\delta\ll1$.
The small-angle assumption is used only to remove lateral-force dependence from the load-transfer expression in \eqref{eq:load_transfer}. 
This simplification avoids an algebraic loop, as lateral force also depends on vertical load. 
The full steering dependence is retained in the main prediction dynamics and lateral-tire-force computation. 
Vertical load is given by
\begin{subequations}
\begin{align} \label{eq:load_transfer}
    F_{zF}&=F_{z,\mathrm{tot}}\frac{l_R}{l_\mathrm{wb}}+\Delta F_z,\\
    F_{zR}&=F_{z,\mathrm{tot}}\frac{l_F}{l_\mathrm{wb}}-\Delta F_z,
\end{align}
\end{subequations}
with
\begin{align}
    F_{z,\mathrm{tot}} &= m(\dot{v}_z + \hat{\phi}' \dot{s} v_y - \hat{\theta}' \dot{s} v_x + g \mathrm{c}_\theta \mathrm{c}_\phi) - F_l, \\
    \Delta F_z &= \frac{1}{l_\mathrm{wb}} ( I_z \dot{\psi} \hat{\phi}' \dot{s} - \Delta h F_d - h (F_{xF} + F_{xR})),
\end{align}
where $\dot{v}_z$ is approximated by $\dot{v}_z=\dot{d}\hat{\phi}'\dot{s}+d\hat{\phi}''\dot{s}^2+d\hat{\phi}'\ddot{s}$ and $\ddot{s}$ is approximated by 
\begin{align} \label{eq:ax_approx_for_vertical_load}
    \ddot{s}\approx
    \frac{1}{m}(T_\mathrm{eff} C_T-B(C_{BF}+C_{BR})-mgC_{rr}-F_d),
\end{align}
which is valid for $\Delta\psi,d\hat{\chi}',\delta \ll 1$.

The coupling of longitudinal and vertical tire loads creates a problematic incentive: when the rear tire sideslip angle saturates, increasing the throttle raises the predicted rear axle vertical load and thus apparent lateral capability. 
While this is plausible for the idealized rigid \ac{dstm} \eqref{eq:dstm_dynamics}, the real vehicle includes a suspension and thus pitch dynamics. 
This means the system reacts more slowly than the prediction model anticipates. 
In practice, the vertical load on the rear axle cannot increase quickly enough to offset the loss in lateral stiffness caused by longitudinal slip. 
Thus, increasing the throttle does not raise rear axle traction but triggers wheel spin and traction control intervention. 
To avoid optimistic load transfer prediction in this situation, we define $T_\mathrm{eff}=T$ in \eqref{eq:ax_approx_for_vertical_load} for body sideslip angle $\beta=\arctan(\frac{v_y}{v_x})\in[-0.06,0.06]\,\si{\radian}$ but set $T_\mathrm{eff}=0$ for sideslip values outside this range.
While including pitch dynamics in the prediction model would be a more \emph{physical} and less heuristic solution, it introduces additional states (body pitch) and parameters (pitch stiffness, inertia). 
Thus, the presented solution with $T_\mathrm{eff}$ targets a pragmatic trade-off between complexity and accuracy. 

\paragraph{Pacejka model and lateral force}
The lateral acceleration is generated by the front and rear axle slip angles $\alpha_i$ given by
\begin{align} \label{eq:slip_angles}
    \alpha_F &=-\arctan\left(\frac{v_y+l_F\dot{\psi}}{v_x}\right)+\delta,\\
    \alpha_R &=-\arctan\left(\frac{v_y-l_R\dot{\psi}}{v_x}\right).
\end{align}
We use a modification of the \ac{mf} tire force law by Pacejka \cite{Pacejka2012TireAndVehicleDynamics} to describe how axle sideslip angle generates lateral force, dependent on vertical and longitudinal load:
%
\begin{align}
      D_i &= (p_{Dy1,i} + p_{Dy2,i} dF_{zi}) \lambda_{\mu,i},\\
      F_{yi} &= F_{zi} D_i G_{y\kappa,i} \sin ( C_i \arctan ( B_i \alpha_i - E_i ( B_i \alpha_i \nonumber\\
      &\quad - \arctan ( B_i \alpha_i ) ) ) ),
\end{align}
where $i \in \{ F, R \}$, and $dF_{zi} = \frac{F_{zi} - F_{z0,i}}{F_{z0,i}}$ with nominal load $F_{z0,i}$.
$G_{y\kappa,i} = \cos(\arctan(C_{Gy,i} F_{xi}))$ couples longitudinal force to lateral dynamics. 
$p_{Dy1,i}$ and $p_{Dy2,i}$ represent the affine dependence on the axle load, and $\lambda_{\mu,i}$ is an overall scaling factor. 
$B_i$, $C_i$ and $E_i$ are empirical shape parameters.

%% file: figures/3d_track.pdf_tex
\begingroup%
  \makeatletter%
  \providecommand\color[2][]{%
    \errmessage{(Inkscape) Color is used for the text in Inkscape, but the package 'color.sty' is not loaded}%
    \renewcommand\color[2][]{}%
  }%
  \providecommand\transparent[1]{%
    \errmessage{(Inkscape) Transparency is used (non-zero) for the text in Inkscape, but the package 'transparent.sty' is not loaded}%
    \renewcommand\transparent[1]{}%
  }%
  \providecommand\rotatebox[2]{#2}%
  \newcommand*\fsize{\dimexpr\f@size pt\relax}%
  \newcommand*\lineheight[1]{\fontsize{\fsize}{#1\fsize}\selectfont}%
  \ifx\svgwidth\undefined%
    \setlength{\unitlength}{363.96498312bp}%
    \ifx\svgscale\undefined%
      \relax%
    \else%
      \setlength{\unitlength}{\unitlength * \real{\svgscale}}%
    \fi%
  \else%
    \setlength{\unitlength}{\svgwidth}%
  \fi%
  \global\let\svgwidth\undefined%
  \global\let\svgscale\undefined%
  \makeatother%
  \begin{picture}(1,0.85027673)%
    \lineheight{1}%
    \setlength\tabcolsep{0pt}%
    \put(0,0){\includegraphics[width=\unitlength,page=1]{3d_track.pdf}}%
    \put(0.21295763,0.42213418){\color[rgb]{0.49019608,0.57254902,0.16470588}\makebox(0,0)[lt]{\lineheight{1.25}\smash{\begin{tabular}[t]{l}$y_\mathcal{T}$\end{tabular}}}}%
    \put(0.14179567,0.48373447){\color[rgb]{0.49019608,0.57254902,0.16470588}\makebox(0,0)[lt]{\lineheight{1.25}\smash{\begin{tabular}[t]{l}$x_\mathcal{T}$\end{tabular}}}}%
    \put(0.21464533,0.58180448){\color[rgb]{0.49019608,0.57254902,0.16470588}\makebox(0,0)[lt]{\lineheight{1.25}\smash{\begin{tabular}[t]{l}$z_\mathcal{T}$\end{tabular}}}}%
    \put(0.38132711,0.62317464){\color[rgb]{0.18823529,0.43921569,0.70196078}\makebox(0,0)[lt]{\lineheight{1.25}\smash{\begin{tabular}[t]{l}$z_\mathcal{V}$\end{tabular}}}}%
    \put(0.40892005,0.47109448){\color[rgb]{0.18823529,0.43921569,0.70196078}\makebox(0,0)[lt]{\lineheight{1.25}\smash{\begin{tabular}[t]{l}$y_\mathcal{V}$\end{tabular}}}}%
    \put(0.38515818,0.54430426){\color[rgb]{0.18823529,0.43921569,0.70196078}\makebox(0,0)[lt]{\lineheight{1.25}\smash{\begin{tabular}[t]{l}$x_\mathcal{V}$\end{tabular}}}}%
    \put(0.12209364,0.42402705){\color[rgb]{0,0,0}\makebox(0,0)[lt]{\lineheight{1.25}\smash{\begin{tabular}[t]{l}$\phi$ \end{tabular}}}}%
    \put(0.11483322,0.53320299){\color[rgb]{0,0,0}\makebox(0,0)[lt]{\lineheight{1.25}\smash{\begin{tabular}[t]{l}$\theta$\end{tabular}}}}%
    \put(0.0853341,0.72271778){\color[rgb]{0,0,0}\makebox(0,0)[lt]{\lineheight{1.25}\smash{\begin{tabular}[t]{l}Planner Trajectory\end{tabular}}}}%
    \put(0.30766156,0.42727691){\color[rgb]{0,0,0}\makebox(0,0)[lt]{\lineheight{1.25}\smash{\begin{tabular}[t]{l}$\Delta\psi$\end{tabular}}}}%
    \put(0.08016181,0.06485904){\color[rgb]{0.89019608,0.44705882,0.13333333}\makebox(0,0)[lt]{\lineheight{1.25}\smash{\begin{tabular}[t]{l}$y_\mathcal{I}$\end{tabular}}}}%
    \put(0.03539121,0.12653836){\color[rgb]{0.89019608,0.44705882,0.13333333}\makebox(0,0)[lt]{\lineheight{1.25}\smash{\begin{tabular}[t]{l}$x_\mathcal{I}$\end{tabular}}}}%
    \put(0.08968284,0.20496585){\color[rgb]{0.89019608,0.44705882,0.13333333}\makebox(0,0)[lt]{\lineheight{1.25}\smash{\begin{tabular}[t]{l}$z_\mathcal{I}$\end{tabular}}}}%
    \put(0.29139576,0.4981629){\color[rgb]{0,0,0}\makebox(0,0)[lt]{\lineheight{1.25}\smash{\begin{tabular}[t]{l}$d$\end{tabular}}}}%
    \put(0.32586212,0.56808215){\color[rgb]{0.60784314,0.2745098,0.55294118}\makebox(0,0)[lt]{\lineheight{1.25}\smash{\begin{tabular}[t]{l}$\bm{v}$\end{tabular}}}}%
    \put(0,0){\includegraphics[width=\unitlength,page=2]{3d_track.pdf}}%
    \put(0.77491212,0.14453742){\color[rgb]{0.60784314,0.2745098,0.55294118}\makebox(0,0)[lt]{\lineheight{1.25}\smash{\begin{tabular}[t]{l}$\bm{v}$\end{tabular}}}}%
    \put(0.61126573,0.23614428){\color[rgb]{0,0,0}\makebox(0,0)[lt]{\lineheight{1.25}\smash{\begin{tabular}[t]{l}Road Plane View:\end{tabular}}}}%
    \put(0.6611189,0.19305305){\color[rgb]{0.18823529,0.43921569,0.70196078}\makebox(0,0)[lt]{\lineheight{1.25}\smash{\begin{tabular}[t]{l}$y_\mathcal{V}$\end{tabular}}}}%
    \put(0.64779044,0.06804623){\color[rgb]{0.18823529,0.43921569,0.70196078}\makebox(0,0)[lt]{\lineheight{1.25}\smash{\begin{tabular}[t]{l}$z_\mathcal{V}$\end{tabular}}}}%
    \put(0.76653278,0.0608904){\color[rgb]{0,0,0}\makebox(0,0)[lt]{\lineheight{1.25}\smash{\begin{tabular}[t]{l}$\beta$\end{tabular}}}}%
    \put(0.81974309,0.1939977){\color[rgb]{0.89019608,0.44705882,0.13333333}\makebox(0,0)[lt]{\lineheight{1.25}\smash{\begin{tabular}[t]{l}$F_{yF}$\end{tabular}}}}%
    \put(0.55779967,0.18914122){\color[rgb]{0.89019608,0.44705882,0.13333333}\makebox(0,0)[lt]{\lineheight{1.25}\smash{\begin{tabular}[t]{l}$F_{yR}$\end{tabular}}}}%
    \put(0.55435976,0.04238198){\color[rgb]{0.49019608,0.57254902,0.16470588}\makebox(0,0)[lt]{\lineheight{1.25}\smash{\begin{tabular}[t]{l}$F_{xR}$\end{tabular}}}}%
    \put(0.93377607,0.15966925){\color[rgb]{0.49019608,0.57254902,0.16470588}\makebox(0,0)[lt]{\lineheight{1.25}\smash{\begin{tabular}[t]{l}$F_{xF}$\end{tabular}}}}%
    \put(0.94833166,0.10555511){\color[rgb]{0,0,0}\makebox(0,0)[lt]{\lineheight{1.25}\smash{\begin{tabular}[t]{l}$\delta$\end{tabular}}}}%
    \put(0.70961764,0.06700161){\color[rgb]{0.18823529,0.43921569,0.70196078}\makebox(0,0)[lt]{\lineheight{1.25}\smash{\begin{tabular}[t]{l}$x_\mathcal{V}$\end{tabular}}}}%
    \put(0.63759417,0.12690182){\color[rgb]{0,0,0}\makebox(0,0)[lt]{\lineheight{1.25}\smash{\begin{tabular}[t]{l}$\dot{\psi}$\end{tabular}}}}%
    \put(0.70164611,0.15194258){\color[rgb]{0.05490196,0.22352941,0.43137255}\makebox(0,0)[lt]{\lineheight{1.25}\smash{\begin{tabular}[t]{l}$mg$\end{tabular}}}}%
    \put(0,0){\includegraphics[width=\unitlength,page=3]{3d_track.pdf}}%
    \put(0.70362222,0.79518144){\color[rgb]{0.18823529,0.43921569,0.70196078}\makebox(0,0)[lt]{\lineheight{1.25}\smash{\begin{tabular}[t]{l}$z_\mathcal{V}$\end{tabular}}}}%
    \put(0.75451434,0.59730339){\color[rgb]{0,0,0}\makebox(0,0)[lt]{\lineheight{1.25}\smash{\begin{tabular}[t]{l}$l_F$\end{tabular}}}}%
    \put(0.60112347,0.56972256){\color[rgb]{0,0,0}\makebox(0,0)[lt]{\lineheight{1.25}\smash{\begin{tabular}[t]{l}$l_R$\end{tabular}}}}%
    \put(0.76305026,0.7478673){\color[rgb]{0.18823529,0.43921569,0.70196078}\makebox(0,0)[lt]{\lineheight{1.25}\smash{\begin{tabular}[t]{l}$x_\mathcal{V}$\end{tabular}}}}%
    \put(0.62297845,0.62690167){\color[rgb]{0,0,0}\makebox(0,0)[lt]{\lineheight{1.25}\smash{\begin{tabular}[t]{l}$\Delta l$\end{tabular}}}}%
    \put(0.71471407,0.67732497){\color[rgb]{0.05490196,0.22352941,0.43137255}\makebox(0,0)[lt]{\lineheight{1.25}\smash{\begin{tabular}[t]{l}$mg$\end{tabular}}}}%
    \put(0.495883,0.69418627){\color[rgb]{0,0,0}\makebox(0,0)[lt]{\lineheight{1.25}\smash{\begin{tabular}[t]{l}$h$\end{tabular}}}}%
    \put(0.64855547,0.82940203){\color[rgb]{0,0,0}\makebox(0,0)[lt]{\lineheight{1.25}\smash{\begin{tabular}[t]{l}Side View:\end{tabular}}}}%
    \put(0.472058,0.63363271){\color[rgb]{0.05490196,0.22352941,0.43137255}\makebox(0,0)[lt]{\lineheight{1.25}\smash{\begin{tabular}[t]{l}$F_{zR}$\end{tabular}}}}%
    \put(0.87259183,0.63747227){\color[rgb]{0.05490196,0.22352941,0.43137255}\makebox(0,0)[lt]{\lineheight{1.25}\smash{\begin{tabular}[t]{l}$F_{zF}$\end{tabular}}}}%
    \put(0.58788182,0.77290681){\color[rgb]{0.05490196,0.22352941,0.43137255}\makebox(0,0)[lt]{\lineheight{1.25}\smash{\begin{tabular}[t]{l}$F_{l}$\end{tabular}}}}%
    \put(0.5755386,0.71523813){\color[rgb]{0.49019608,0.57254902,0.16470588}\makebox(0,0)[lt]{\lineheight{1.25}\smash{\begin{tabular}[t]{l}$F_{d}$\end{tabular}}}}%
    \put(0.47022918,0.73359993){\color[rgb]{0,0,0}\makebox(0,0)[lt]{\lineheight{1.25}\smash{\begin{tabular}[t]{l}$\Delta h$\end{tabular}}}}%
    \put(0.70354634,0.75271906){\color[rgb]{0.18823529,0.43921569,0.70196078}\makebox(0,0)[lt]{\lineheight{1.25}\smash{\begin{tabular}[t]{l}$y_\mathcal{V}$\end{tabular}}}}%
    \put(0,0){\includegraphics[width=\unitlength,page=4]{3d_track.pdf}}%
  \end{picture}%
\endgroup%

%% file: sections/control_design.tex
\subsection{Control Design}

\subsubsection{MPC Setup}

The \ac{nlp} solved at each controller update step is given in \eqref{eq:mpc_ocp}. 
It is similar to \cite{Bongard2025IVRMPC}, with a cost that balances tracking and time optimality, prioritizes vehicle stability, and accounts for unmodeled dynamics.

\begin{subequations} \label{eq:mpc_ocp}
	\begin{align}
    \MoveEqLeft[2]
    \min_{\substack{\bm{x}_0, \dots, \bm{x}_N\\ \bm{u}_0, \dots, \bm{u}_{N-1}\\ \bm{\mathcal{s}}_0, \dots, \bm{\mathcal{s}}_{N-1}}} \sum_{k=0}^{N-1} \mathbf{y}_k^\top \mathbf{Q} \mathbf{y}_k + \mathbf{u}_k^\top \mathbf{R} \mathbf{u}_k + \rho(\bm{\mathcal{s}}_k) + \mathbf{y}_N^\top \mathbf{Q}_N \mathbf{y}_N \label{eq:mpc_cost_overall}\\
		\text{s.\,t.}\ &\mathbf{x}_{k+1} = \mathbf{f}_d(\mathbf{x}_k, \mathbf{u}_k, \mathbf{p}_k), \label{eq:mpc_dynamics}\\
		&d_{\mathrm{lb}} - \mathcal{s}_k^1  \leq d_k \leq d_{\mathrm{ub}} + \mathcal{s}_k^2, \label{eq:constr_d}\\
		&\delta_\mathrm{lb} - \mathcal{s}_k^3 \leq \delta_k \leq \delta_\mathrm{ub} + \mathcal{s}_k^4, \label{eq:constr_delta}\\
		&0 \leq T_k \leq 1 + \mathcal{s}_k^5, \quad 0 \leq B_k \leq 1 + \mathcal{s}_k^6, \label{eq:constr_TandB}\\
		&\mathbf{u}_{\mathrm{lb}} - \bm{\mathcal{s}}_k^7 \leq \mathbf{u}_k \leq \mathbf{u}_{\mathrm{ub}} + \bm{\mathcal{s}}_k^8, \label{eq:constr_u}\\
		&\tilde{a}_{x,k} + m_{r,k} \tilde{a}_{y,k} \leq C_{r,k} + \mathcal{s}_k^{9,r} \quad r \in \mathbb{I}_{[1,\,8]}, \label{eq:constr_gg_diagram}\\
		&\alpha_{i,\mathrm{lb}} - \mathcal{s}_k^{10,i} \leq \alpha_{i,k} \leq \alpha_{i,\mathrm{ub}} + \mathcal{s}_k^{11,i} \quad i \in \{ F, R \},\label{eq:constr_slip_angles}\\
        &\bm{0} \leq \bm{\mathcal{s}}_k, \label{eq:constr_slacks}\\
		&v_{N} \leq v_{\mathrm{ref}, N} \label{eq:constr_v_terminal}
	\end{align}
\end{subequations}

The constraints \eqref{eq:mpc_dynamics} to \eqref{eq:constr_slacks} are imposed for $k \in \mathbb{I}_{\left[ 0, N-1 \right]}$.

The cost terms in \eqref{eq:mpc_cost_overall} are given by
\begin{subequations}
\begin{align} 
	\mathbf{y}_k &= \begin{bmatrix}
		d_k,
		\dot{d}_k,
		\Delta v_k,
		T_k B_k,
		\dot{a}_{y,\mathrm{kin},k},
		\frac{v_k}{\dot{s}_k}
	\end{bmatrix}^\top,  \label{eq:mpc_stage_cost}\\
	\rho(\bm{\mathcal{s}}_k) &= \sum_{i=1}^{n_{\bm{\mathcal{s}}}} q_{1}^i \mathcal{s}_k^i + q_{2}^i {\mathcal{s}_k^i}^2. \label{eq:mpc_slack_cost}\\
    \mathbf{y}_N &= [\dot{d}_N]. \label{eq:mpc_terminal_cost}
\end{align}
\end{subequations}

The \ac{nlp} \eqref{eq:mpc_ocp} uses reference trajectory information $\bm{p}_k$ comprised of velocity $v_\mathrm{ref,k}$, banking, slope, curvature, and acceleration limits (in the form of $m_{r,k}$, $C_{r,k}$ as in \eqref{eq:constr_gg_diagram}) provided by the higher-level planning module. The reference trajectory signals are treated as purely time-varying under the assumption that the reference is tracked closely enough.

The stage cost \eqref{eq:mpc_stage_cost} is chosen such that $d$, $\dot{d}$, and $\Delta v = v_k - v_\mathrm{ref,k}$ incentivize tracking of the reference path and velocity. 
By including $\dot{d}$, we penalize lateral motion relative to the reference line, not just the absolute offset $d$.
As a result, the controller is incentivized to minimize heading errors, especially at higher speeds, which improves tracking performance and stability.
The product $TB$ penalizes simultaneous braking and throttle application, which is possible in the model but not intended in our overall hierarchical controller concept. 
Adding costs on the kinematic jerk $\dot{a}_{y,\mathrm{kin}} = \frac{d}{dt} \left( \delta v_x^2 \right) \approx \dot{\delta} v_{x}^2 + 2 \delta \tilde{a}_x v_x$ incentivizes smooth lateral acceleration especially at higher velocities which helps to account for unmodeled dynamics.

The terminal cost term in \eqref{eq:mpc_terminal_cost} penalizes lateral path error velocity at the end of the horizon. 
By driving $\dot{d}_N \rightarrow 0$, the predicted terminal state approaches motion parallel to the longer-term planner trajectory.
Thus, we improve recursive feasibility for the short-term \ac{mpc} prediction horizon.

For computational efficiency, the coupled acceleration envelope is represented as an octagon in the apparent-acceleration plane.
Compared to an elliptic formulation, this yields eight linear inequalities in $(\tilde a_x,\tilde a_y)$ and simpler derivatives, while also being compatible with the message interface between planning and control modules.
Note that the resulting constraints \eqref{eq:constr_gg_diagram} are still nonlinear in the optimization variables through the mapping $(\bm{x},\bm{u}) \mapsto (\tilde a_x,\tilde a_y)$.

The axle slip angle constraints \eqref{eq:constr_slip_angles} are imposed as additional stability constraints. They discourage excessive tire sideslip angles by restricting the controller from using the nonlinear tire curve beyond its lateral traction force peak. 

The constraints \eqref{eq:constr_slacks} on the slacks $\bm{\mathcal{s}}_k \in \R^{n_{\bm{\mathcal{s}}}}$ represent element-wise non-negativity constraints ensuring slacks only relax selected constraints, which improves \ac{nlp} feasibility under unforeseen circumstances.
Meanwhile, linear and quadratic costs \eqref{eq:mpc_slack_cost} on the slacks keep the \ac{nlp} close to the constrained solution when it is feasible.

The terminal constraint \eqref{eq:constr_v_terminal} incentivizes keeping below the terminal target velocity, intended for recursive feasibility and safety.

\subsubsection{Constraint Tightening (CT)}

In autonomous racing, it is critical not to stress tires beyond their capabilities of longitudinal and lateral traction. 
Exceeding these limits results in excessive slip, a drop-off of tire forces, and ultimately a loss of vehicle stability in the form of critical over- or under-steer.
Stability constraints in the form of coupled acceleration bounds at the center of mass can be added to the optimization problem as a proxy for tire friction potential \cite{Stano2023}.
However, a nominal \ac{mpc} with stability constraints does not consider that real-world vehicle behavior, especially tire-road interaction, is inherently uncertain. 
Robust tube-\ac{mpc} approaches extend nominal \ac{mpc} by considering the propagation of uncertainty through the prediction model and introducing a margin to constraint bounds, with the aim of keeping the vehicle controllable and stable.
In the following, we recall the previously introduced dynamic \ac{ct} approach based on \ac{ccm} analysis for a two-dimensional single-track model under bounded uncertainty \cite{Bongard2025IVRMPC}.

The constraints \eqref{eq:constr_d} to \eqref{eq:constr_gg_diagram} as well as the terminal speed constraint \eqref{eq:constr_v_terminal} are formulated as one-sided constraints of the form $h_j(\bm{x}, \bm{u}) \leq 0$ (ignoring slack variables for notational ease).
\acf{ct} is applied with a time-dependent scalar \emph{tube size} $\sigma(t)$ and corresponding tightening constant $c_j$
\begin{align}
	 h_j(\bm{x}, \bm{u}) + c_j \sigma \leq 0. \label{eq:constraint_tightening}
\end{align}
The dynamics of the tube size $\sigma$ is then given as a function of nominal state and input
\begin{align} \label{eq:tube_dynamics}
	\dot{\sigma} = - \left(\beta - L_\mathbb{E} - \mathcal{C}_\sigma \right) \sigma + f_\sigma(\bm{x}, \bm{u}).
\end{align}
Thus, the tube size $\sigma$, characterizing our uncertainty around the nominal state and input, effectively becomes a state variable in the \ac{mpc} prediction model.
For notational clarity, we keep the robustification scheme with the tube dynamics \eqref{eq:tube_dynamics} separate from our nominal \ac{nlp} defined in \eqref{eq:mpc_ocp}.

In the tube dynamics \eqref{eq:tube_dynamics}, $\beta$ and $L_\mathbb{E}$ are constants related to the achievable contraction of the uncertain model under external disturbances. $\mathcal{C}_\sigma$ is a constant related to the unknown but bounded parameter error vector.
The state- and input-dependent \emph{uncertainty} $f_\sigma(\bm{x}, \bm{u})$ is a worst-case bound over the vertices of the uncertain but bounded parameter and external disturbance vertices according to the \ac{ccm} analysis \cite{Bongard2025IVRMPC}. By design, $f_\sigma$ increases with the states relevant to vehicle stability: $v_x$, $v_y$, $\dot{\psi}$, and the longitudinal forces via $T$ and $B$. These states drive the uncertainty dynamics \eqref{eq:tube_dynamics}.
We refer to \cite{Bongard2025IVRMPC} for the computation of these expressions, which is performed offline.


The tightening of path deviation constraints \eqref{eq:constr_d} and acceleration constraints \eqref{eq:constr_gg_diagram} leads to a safety distance from path bounds and withheld acceleration potential later in the \ac{mpc} horizon, respectively. 
The coupling of constraint and nominal dynamics \eqref{eq:constraint_tightening} allows the controller to consider uncertainties within the optimization. 
This enables the controller to automatically balance the tradeoff between trajectory tracking and vehicle stability \cite{Bongard2025IVRMPC}.

The proposed \ac{ct} \eqref{eq:constraint_tightening}, \eqref{eq:tube_dynamics} is intended to assist in keeping problem \eqref{eq:mpc_ocp} recursively feasible under uncertainty by covering the assumed set of uncertain trajectories within an uncertainty-dependent \emph{tube}, and using this tube as a back-off from constraint bounds \cite{Sasfi2022}. The \ac{ct} is dependent on the uncertainty in order to tightly cover the set of uncertain trajectories and therefore induce the right amount of caution while avoiding unnecessary conservatism. 
In practice, the tube dynamics \eqref{eq:tube_dynamics} grow during aggressive maneuvers and tend to contract as the vehicle returns to nominal operating conditions. The limited number and interpretability of constants in \eqref{eq:constraint_tightening} make the \ac{ct} tuneable by hand.

\begin{figure}[t!]
    \centering
    \begin{subfigure}[t]{0.48\columnwidth}
        \centering
        \includegraphics[width=\linewidth]{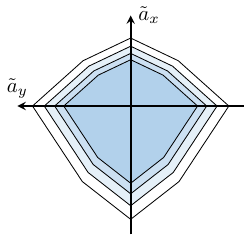}
    \end{subfigure}%
    \hfill
    \begin{subfigure}[t]{0.48\columnwidth}
        \centering
        \includegraphics[width=\linewidth]{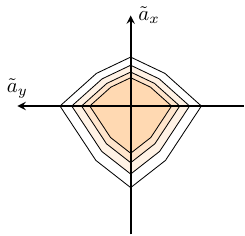}
    \end{subfigure}
    \caption{Side-by-side illustration of progressive tightening of acceleration constraints \eqref{eq:constr_gg_diagram} along the \ac{mpc} horizon for two different (static) regimes of apparent gravity due to \ac{3d} racetrack geometry: Higher apparent gravity on the left, lower on the right. Shading indicates the horizon index, with low saturation at the start and high saturation at the end.}
    \label{fig:tightened_octagon}
\end{figure}

The \ac{mpc} stability constraints \eqref{eq:constr_gg_diagram} on the apparent accelerations \eqref{eq:accels_tilde}, provided by the planning module, are strongly influenced by \ac{3d} racetrack geometry \cite{Rowold2023}. 
While the \ac{ct} leads to a more conservative behavior, the withheld acceleration potential is invariant under vertical loads. Instead, \ac{3d} racetrack geometry is indirectly considered through its impact on acceleration limits (see Fig.~\ref{fig:tightened_octagon}).

%% file: sections/results.tex
\section{Results}
\begin{figure}
    \centering
    \def\svgwidth{\linewidth} 
    \includegraphics{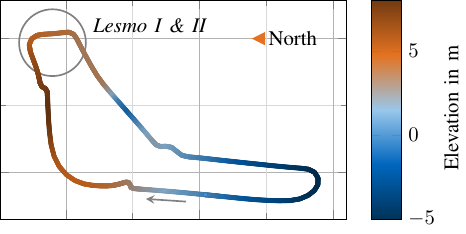}
    \caption{Layout of the \emph{Autodromo Nazionale di Monza} circuit overlaid on a \SI{500}{\m} grid. 
    Track elevation is represented by a color gradient. 
    The curves \emph{Lesmo I \& II}, used later, are highlighted.}
    \label{fig:track}
\end{figure}

\begin{figure}
    \centering
    \def\svgwidth{\linewidth}
    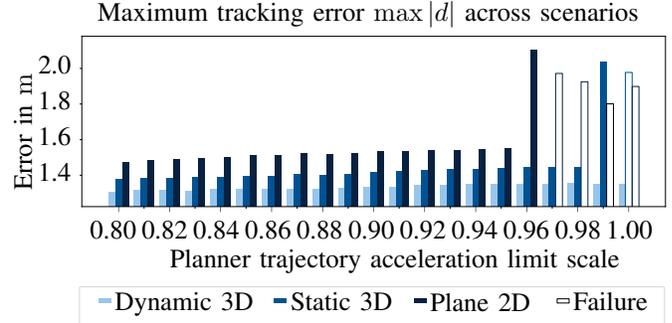
    \caption{Maximum absolute path deviation across two laps driven at various levels of planner trajectory aggressiveness, specified by the scaling value of the acceleration limits.}
    \label{fig:max_abs_d}
\end{figure}

\begin{figure}
    \begin{flushright}
    \def\svgwidth{\linewidth}
	\includegraphics{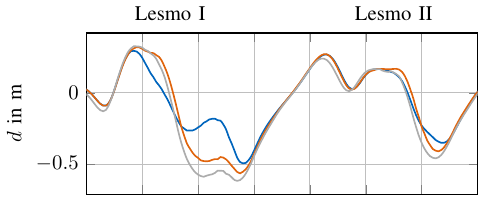}
	\includegraphics{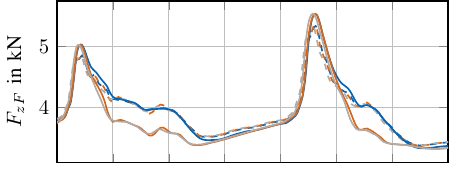}
	\includegraphics{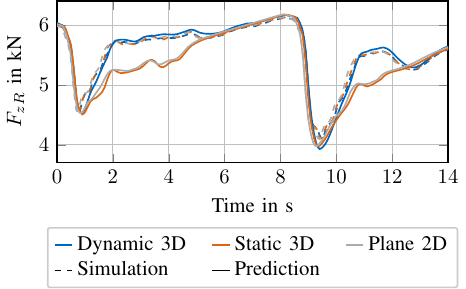}
    \end{flushright}
    \caption{Path deviation $d$ and vertical loads $F_{zF}$, $F_{zR}$ in the turns \emph{Lesmo I} and \emph{II} for the scenario at $0.98$ planner acceleration limit scale. 
    The solid lines in the second and third plots show the loads assumed by the MPC prediction model. 
    The dashed lines show the simulation ground-truth axle load values, calculated as the sum of the left and right tire forces.}
    \label{fig:deviation_normal_forces}
\end{figure}

For real-time capability, the presented controller is implemented with \texttt{C}-Code generated by the \texttt{acados} \ac{mpc} framework \cite{Verschueren2021acados}.
The \texttt{C}-Code is integrated within a \texttt{C++} node for the ROS2-based software stack used by \emph{TUM Autonomous Motorsport}~\cite{Betz2019SoftwareArchitecture,hoffmann2025a2rl}. 
In this architecture, the reference trajectory with \ac{3d} geometry information and constraints is generated by a sampling-based planning algorithm~\cite{gretmen2024}.
The prediction model \eqref{eq:mpc_dynamics} is discretized using a Runge-Kutta method of order $4$ with a step size of \SI{60}{\ms} and a horizon length of $36$ steps, giving a prediction horizon of \SI{2.16}{\s}.
To minimize latency, we use the \ac{rti} scheme \cite{Diehl2002RTI}: for each control update, the \ac{nlp} \eqref{eq:mpc_ocp} is linearized once around a warm start from the previous solution, one QP subproblem is solved, and the control input is extracted from the result. 
We employ \texttt{HPIPM}~\cite{Frison2020} with a fixed number of iterations for solving the resulting \ac{qp} in the intermediate speed setting \texttt{BALANCE}. 
The controller is running at an update frequency of \SI{100}{\hertz}.

The normalized throttle and brake forces optimized in \ac{nlp} \eqref{eq:mpc_ocp} are converted to an acceleration request for the low-level longitudinal controller~\cite{Pitschi2025}. 
Conversely, the measured longitudinal acceleration is approximately converted to normalized throttle and brake forces as acceleration feedback.

To test the impact of the proposed changes, we evaluate three controller variants on a set of $20$ scenarios in a high-fidelity dynamic double-track vehicle simulation \cite{Sagmeister2024}. 
The \emph{Plane \ac{2d}} variant neglects all \ac{3d} effects. 
The \emph{Static \ac{3d}} variant considers static banking and slope effects as in \cite{Raji2022}.
The \emph{Dynamic \ac{3d}} variant considers all \ac{3d} acceleration outlined in \ref{sec:vehicle_model}. 
All variants are parameterized with the same set of cost and slack weights.

Figure~\ref{fig:track} shows the \emph{Autodromo Nazionale di Monza} racetrack, which is used for all $20$ evaluation scenarios. 
The track's unique topology, with an elevation change of \SI{13}{\m}, a maximum slope of $2.6\%$, and speeds up to \SI{255}{\km\per\hour}, makes it a suitable testing ground for investigating the influence of dynamic acceleration. 
In this environment, vertical acceleration changes are mainly induced by the rotation of the vehicle traveling at high velocity on a curved road surface, rather than the absolute orientation (i.e., banking and slope). 
Each scenario consists of two laps, one out-lap and one race lap. 
The different scenarios are characterized by the scaling of the acceleration constraints used to generate the planner trajectories and the \ac{mpc} acceleration limits \eqref{eq:constr_gg_diagram}.

Figure~\ref{fig:max_abs_d} demonstrates the benefit of using the \ac{3d}-\ac{dstm} in the \ac{mpc} across the different scenarios. 
The proposed control algorithm with the \ac{3d}-\ac{dstm} outperforms the plane \ac{2d} and static \ac{3d} variants by giving lower maximum absolute path errors. 
The \emph{Dynamic \ac{3d}} variant is also able to stabilize the vehicle at higher acceleration limits.

To illustrate how the lower tracking error can be achieved with the \emph{Dynamic \ac{3d}} variant, we plot the lateral error and predicted vertical tire load in the corners \emph{Lesmo I} and \emph{II} (Figure \ref{fig:track}).
In the first corner (\emph{Lesmo I}), the racetrack slopes down and then levels for the apex of the corner. 
As a result of this change in slope, the ground-truth simulation vertical forces on the front and rear axles are significantly higher compared to the predictions of the \emph{Plane 2D} and \emph{Static \ac{3d}} variants (Figure \ref{fig:deviation_normal_forces} center and bottom). 
The \emph{Dynamic \ac{3d}} approach with a rotating reference frame captures terrain dynamics more accurately, resulting in tire loads closer to the ground-truth values. 
Thus, the \emph{Dynamic \ac{3d}} approach ultimately achieves superior tracking-performance on non-planar racetrack pieces (Figure \ref{fig:deviation_normal_forces} top).
A similar effect can be observed to a lesser extent in the second corner (\emph{Lesmo II}).

\begin{figure}
    \centering
    \def\svgwidth{\linewidth}
    \includegraphics{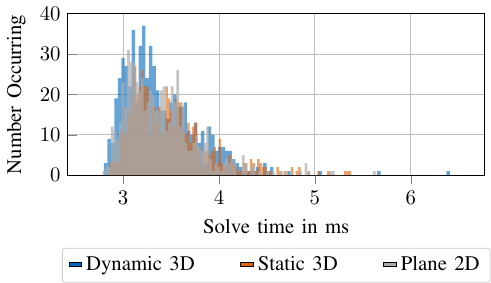}
    \caption{Overlaid histograms of MPC solve times for different prediction model levels of detail.}
    \label{fig:solve_times_histogram}
\end{figure}

Figure \ref{fig:solve_times_histogram} compares the distribution of solve times for the three controller variants. 
The tests were performed on an AMD Ryzen 7 PRO 7840U CPU with AVX2/AVX-512 support, rated to \SI{2.7}{\giga\hertz}. 
Across \SI{60}{\s} of runtime, we observe a similar average solve time of approximately \SI{3.4}{\milli\s} for all three variants.
Peak solve time is slightly higher in the \emph{Dynamic \ac{3d}} variant compared to \SI{5.3}{\milli\s} in the \emph{Static \ac{3d}} and \SI{5.6}{\milli\s} in the \emph{Plane \ac{2d}} variants.
The solve time median is \SI{3.3}{\milli\s} for the \emph{Dynamic \ac{3d}}, \SI{3.4}{\milli\s} for the \emph{Static \ac{3d}}, and \SI{3.3}{\milli\s} for the \emph{Plane \ac{2d}} variants.
The distribution of solve times for all three variants indicates that the solve times remain feasible.

%% file: figures/rslcpp_gg_runthrough_monza_max_abs_d.pdf_tex
\begingroup%
  \makeatletter%
  \providecommand\color[2][]{%
    \errmessage{(Inkscape) Color is used for the text in Inkscape, but the package 'color.sty' is not loaded}%
    \renewcommand\color[2][]{}%
  }%
  \providecommand\transparent[1]{%
    \errmessage{(Inkscape) Transparency is used (non-zero) for the text in Inkscape, but the package 'transparent.sty' is not loaded}%
    \renewcommand\transparent[1]{}%
  }%
  \providecommand\rotatebox[2]{#2}%
  \newcommand*\fsize{\dimexpr\f@size pt\relax}%
  \newcommand*\lineheight[1]{\fontsize{\fsize}{#1\fsize}\selectfont}%
  \ifx\svgwidth\undefined%
    \setlength{\unitlength}{1084.64733887bp}%
    \ifx\svgscale\undefined%
      \relax%
    \else%
      \setlength{\unitlength}{\unitlength * \real{\svgscale}}%
    \fi%
  \else%
    \setlength{\unitlength}{\svgwidth}%
  \fi%
  \global\let\svgwidth\undefined%
  \global\let\svgscale\undefined%
  \makeatother%
  \begin{picture}(1,0.50310007)%
    \lineheight{1}%
    \setlength\tabcolsep{0pt}%
    \put(0,0){\includegraphics[width=\unitlength,page=1]{rslcpp_gg_runthrough_monza_max_abs_d.pdf}}%
    \put(0.03393153,0.38664313){\color[rgb]{0,0,0}\makebox(0,0)[lt]{\lineheight{1.25}\smash{\begin{tabular}[t]{l}2.0\end{tabular}}}}%
    \put(0.03367583,0.33187277){\color[rgb]{0,0,0}\makebox(0,0)[lt]{\lineheight{1.25}\smash{\begin{tabular}[t]{l}1.8\end{tabular}}}}%
    \put(0.03363622,0.27710243){\color[rgb]{0,0,0}\makebox(0,0)[lt]{\lineheight{1.25}\smash{\begin{tabular}[t]{l}1.6\end{tabular}}}}%
    \put(0.0335858,0.22232486){\color[rgb]{0,0,0}\makebox(0,0)[lt]{\lineheight{1.25}\smash{\begin{tabular}[t]{l}1.4\end{tabular}}}}%
    \put(0.02451654,0.207299){\color[rgb]{0,0,0}\rotatebox{90}{\makebox(0,0)[lt]{\lineheight{1.25}\smash{\begin{tabular}[t]{l}Error in $\mathrm{m}$\end{tabular}}}}}%
    \put(0,0){\includegraphics[width=\unitlength,page=2]{rslcpp_gg_runthrough_monza_max_abs_d.pdf}}%
    \put(0.12579499,0.46849753){\color[rgb]{0,0,0}\makebox(0,0)[lt]{\lineheight{1.25}\smash{\begin{tabular}[t]{l}Maximum tracking error $\max|d|$ across scenarios\end{tabular}}}}%
    \put(0,0){\includegraphics[width=\unitlength,page=3]{rslcpp_gg_runthrough_monza_max_abs_d.pdf}}%
    \put(0.15223385,0.02241212){\color[rgb]{0,0,0}\makebox(0,0)[lt]{\lineheight{1.25}\smash{\begin{tabular}[t]{l}Dynamic 3D\end{tabular}}}}%
    \put(0.42264859,0.0219976){\color[rgb]{0,0,0}\makebox(0,0)[lt]{\lineheight{1.25}\smash{\begin{tabular}[t]{l}Static 3D\end{tabular}}}}%
    \put(0.63753243,0.02098236){\color[rgb]{0,0,0}\makebox(0,0)[lt]{\lineheight{1.25}\smash{\begin{tabular}[t]{l}Plane 2D\end{tabular}}}}%
    \put(0,0){\includegraphics[width=\unitlength,page=4]{rslcpp_gg_runthrough_monza_max_abs_d.pdf}}%
    \put(0.85768371,0.02251521){\color[rgb]{0,0,0}\makebox(0,0)[lt]{\lineheight{1.25}\smash{\begin{tabular}[t]{l}Failure\end{tabular}}}}%
    \put(0.11376821,0.13304651){\color[rgb]{0,0,0}\makebox(0,0)[lt]{\lineheight{1.25}\smash{\begin{tabular}[t]{l}0.80\end{tabular}}}}%
    \put(0.19329103,0.13304651){\color[rgb]{0,0,0}\makebox(0,0)[lt]{\lineheight{1.25}\smash{\begin{tabular}[t]{l}0.82\end{tabular}}}}%
    \put(0.27281391,0.13304651){\color[rgb]{0,0,0}\makebox(0,0)[lt]{\lineheight{1.25}\smash{\begin{tabular}[t]{l}0.84\end{tabular}}}}%
    \put(0.35233674,0.13304651){\color[rgb]{0,0,0}\makebox(0,0)[lt]{\lineheight{1.25}\smash{\begin{tabular}[t]{l}0.86\end{tabular}}}}%
    \put(0.43185956,0.13304651){\color[rgb]{0,0,0}\makebox(0,0)[lt]{\lineheight{1.25}\smash{\begin{tabular}[t]{l}0.88\end{tabular}}}}%
    \put(0.5113825,0.13304651){\color[rgb]{0,0,0}\makebox(0,0)[lt]{\lineheight{1.25}\smash{\begin{tabular}[t]{l}0.90\end{tabular}}}}%
    \put(0.59090526,0.13304651){\color[rgb]{0,0,0}\makebox(0,0)[lt]{\lineheight{1.25}\smash{\begin{tabular}[t]{l}0.92\end{tabular}}}}%
    \put(0.67042814,0.13304651){\color[rgb]{0,0,0}\makebox(0,0)[lt]{\lineheight{1.25}\smash{\begin{tabular}[t]{l}0.94\end{tabular}}}}%
    \put(0.74995097,0.13304651){\color[rgb]{0,0,0}\makebox(0,0)[lt]{\lineheight{1.25}\smash{\begin{tabular}[t]{l}0.96\end{tabular}}}}%
    \put(0.82947379,0.13304651){\color[rgb]{0,0,0}\makebox(0,0)[lt]{\lineheight{1.25}\smash{\begin{tabular}[t]{l}0.98\end{tabular}}}}%
    \put(0.90834848,0.13304651){\color[rgb]{0,0,0}\makebox(0,0)[lt]{\lineheight{1.25}\smash{\begin{tabular}[t]{l}1.00\end{tabular}}}}%
    \put(0.23669727,0.08914677){\color[rgb]{0,0,0}\makebox(0,0)[lt]{\lineheight{1.25}\smash{\begin{tabular}[t]{l}Planner trajectory acceleration limit scale\end{tabular}}}}%
  \end{picture}%
\endgroup%

%% file: sections/discussion.tex
\section{Discussion}
The presented \ac{mpc} scheme addresses two prominent problems in trajectory-tracking control for autonomous racing, especially on highly \ac{3d} racetracks. 
\\
The first is that performance is heavily dependent on the predictive quality of the plant model in the controller. 
This problem is ameliorated by incorporating the dominant forces from racetrack geometry directly into the prediction model. 
By thus improving the prediction quality (Fig.~\ref{fig:deviation_normal_forces}), the \ac{mpc} is able to exploit road topography to increase tracking performance and vehicle stability (Fig.~\ref{fig:max_abs_d}) while keeping solving times feasible (Fig.~\ref{fig:solve_times_histogram}).\\
The second problem is addressing the inevitable model mismatch under uncertainty and constraints, which can lead to constraint violations and, consequently, crashes. 
The presented \ac{ct} provides an uncertainty-aware dynamic safety margin around constraints and reacts to varying overall normal forces via acceleration limits. 
However, the design is inherently \ac{2d}, and does not incorporate dynamic load changes. 
This simplification enables a fast update rate while still allowing the controller to incorporate dynamic, state-dependent uncertainty into the prediction.\\
The consideration of load transfer effects touches on both problems. 
Longitudinal load transfer significantly influences the tire forces and is thus used by human race drivers \cite{Velenis2007}. 
Its consideration is beneficial for control performance, as it improves the prediction of vertical load and, consequently, tire-ground dynamics. 
However, in edge cases, it can introduce adverse incentives, such as increasing throttle near the oversteer threshold. 
A more detailed coupled lateral/longitudinal tire force model~\cite{Weber2024b} with pitch dynamics could reduce this incentive, but would also be more computationally complex.\\
The presented controller is developed for racing in the context of the \emph{TUM Autonomous Motorsport} project~\cite{hoffmann2025a2rl}.